# Controlled chemical functionalization toward 3D-2D carbon nanohorn-MoS$_2$ heterostructures with enhanced electrocatalytic activity for protons reduction

*Antonia Kagkoura,[1] Raul Arenal,[2,3,4] Nikos Tagmatarchis[1*]*

[1] Theoretical and Physical Chemistry Institute, National Hellenic Research Foundation, 48 Vassileos Constantinou Avenue, 11635 Athens, Greece

[2] Laboratorio de Microscopias Avanzadas (LMA), Universidad de Zaragoza, Mariano Esquillor s/n, 50018 Zaragoza, Spain

[3] Instituto de Nanociencia y Materiales de Aragon (INMA), CSIC-U. de Zaragoza, Calle Pedro Cerbuna 12, 50009 Zaragoza, Spain

[4] ARAID Foundation, 50018 Zaragoza, Spain

E-mail: tagmatar@eie.gr

**Abstract**: The realization of novel heterostructures arising from the combination of nanomaterials is an effective way to modify their physicochemical and electrocatalytic properties, giving them enhanced characteristics stemming from their individual constituents. Interfacing carbon nanohorns (CNHs) possessing high porosity, large specific surface area and good electrical conductivity, with MoS$_2$ owning multiple electrocatalytic active sites but lacking significant conductivity, robust interactions and effective structure, can be a strategy to boost the electrocatalytic reduction of protons to molecular hydrogen. Herein, we covalently introduce, in a stepwise approach, complementary functional groups at the conical tips and sidewalls of CNHs, along with the basal plane of $MoS_2$, en route the construction of 3D-2D CNH-$MoS_2$ heterostructures. The increased $MoS_2$ loading onto CNHs, improving and facilitating charge delocalization and transfer in neighboring CNHs, along with the plethora of active sites, results in excellent electrocatalytic activity for protons reduction same to that of commercial Pt/C. We have registered minute overpotential, low Tafel slope and small charge-transfer resistance for electrocatalyzing the evolution of hydrogen from the newly prepared heterostructure of 0.029 V, 71 mV/dec and 34.5 Ω, respectively. Furthermore, the stability of the 3D-2D CNH-$MoS_2$ heterostructure was validated after performing 10,000 ongoing electrocatalytic cycles.

## 1. Introduction

Carbon nanohorns (CNHs) belong to the wider $sp^2$ hybridized carbon nanostructures family, closely related to carbon nanotubes, though with important structural and physical properties differences. To name the most important ones, (a) the tips of CNHs are conical due to the presence of five-membered rings, (b) CNHs aggregate in 3D spherical superstructures of around 100 nm in diameter, (c) they are metal-free produced, and (d) they contain external micropores and internal mesopores.[1] Particularly, the high porosity, large specific surface area, high thermal and chemical stability, good electrical conductivity of CNHs account for their efficient participation in electrocatalytic reactions, mainly, but not only, as support substrate.[2] Conversely, although in recent years the inherent insolubility of CNHs in wet media has been overcome thanks to the establishment of diverse functionalization protocols for directing the covalent anchorage of organic species at the tips or sidewalls of CNHs,[3-8] facilitating their manipulation, hybridization with other nanostructures is scarce. For example, interfacing CNHs with transition metal dichalcogenides possessing multiple electrocatalytic active sites but lacking significant conductivity, robust interactions and effective structure, toward the formation of novel heterostructures potentially suitable to electrocatalyze, among other transformations, protons reduction to molecular hydrogen, e.g. hydrogen evolution reaction (HER), is restricted.[9]

Bonding arrangement, e.g. surface configuration, and size dimension, e.g. reduced thickness from bulk to 2D nanosheets, govern and significantly affect the electronic structure of $MoS_2$ as benchmark transition metal dichalcogenide material. Although $MoS_2$ is found in two polymorphs, the 2H semiconducting polytype with trigonal prismatic coordinated Mo atoms, and the 1T metallic polymorph with octahedral coordination at Mo, the latter one shows intrinsic reduced charge-transfer resistance making it promising for electrocatalysis.[10] Additionally, unsaturated Mo atoms at the basal plane of 1T-$MoS_2$ are electrocatalytically active, opposed to 2H-$MoS_2$ where catalytically active sites at the basal plane are absent.[11] In this context, 1T-$MoS_2$ has proven to be active in electrocatalyzing protons reduction, especially when hybridized with other species.[12-14] The latter can be achieved by either in-situ preparation and growth of 1T-$MoS_2$ on substrates[12,13] or by chemical modification of 1T-$MoS_2$.[15] Notably, functionalization of 1T-$MoS_2$ does not rely on commonly introduced edge-located sulfur vacancies,[16,17] but mostly rests on electron-rich sulfur species at the basal plane,[18,19] derived from extended electron-transfer during lithium intercalation upon BuLi exfoliation of the bulk material. Thus, substitution reactions with organic halides and aryl diazonium salts facilitate the incorporation of organic addends onto the basal plane of 1T-$MoS_2$.[20-22]

The combination of different nanostructured materials, forming novel heterostructures, inevitably alters the electronic structure and surface of the interfacing materials and allows overcoming limitations of the individual species, while benefitting from the unique characteristics of the different components. Particularly for heterostructures based on atomically thin 2D nanomaterials featuring unprotected and exposed electrons, noteworthy phenomena such as charge-delocalization and transfer between adjacent nanomaterials, band alignment and enhanced redox properties, occur at the heterointerface. Furthermore, when it comes to heterostructures of CNHs with $MoS_2$, the unique structure of the latter, containing electropositive and electronegative elements, introduces properties, which are generally absent in the former nanostructure, containing only carbon in their framework.

The conceptualization of CNHs, due to their 3D porous network with large surface area, excellent electrical conductivity and high chemical stability, to interface 2D $MoS_2$ nanosheets with metallic phase, for electrocatalyzing protons reduction en route hydrogen generation, is of paramount importance and has yet to be realized. To further boost and advance the electrocatalytic properties of such 3D-2D heterostructures, conjugation of increased amount of 1T-$MoS_2$, carrying active catalytic centers for HER, onto CNHs is required. Herein, by employing advanced functionalization methodologies, we initially covalently attached $MoS_2$ nanosheets at the tips of CNHs and then further incorporated additional $MoS_2$ at the sidewalls of CNHs, to realize robust CNH-$MoS_2$ with high $MoS_2$ loading. We carefully followed every modification step and verified the success of performed reactions by FT-IR and Raman spectroscopy together with thermogravimetry and electron microscopy imaging means. As proof-of-concept, the electrocatalytic activity of the 3D-2D heterostructured CNH-$MoS_2$ nanoconjugate toward protons reduction was assessed by linear sweep voltammetry and electrochemical impedance spectroscopy and found that the increased $MoS_2$ loading onto CNHs benefits the process. Markedly, the covalent linkage between CNHs and $MoS_2$ improves charge-transfer to the catalytic sites between heterointerfaces and delivers greater interfacial surface area enhancing the redox reaction. The superior electrocatalytic activity of CNH-$MoS_2$, reaching that of commercial Pt/C, while exhibiting excellent stability and durability, highlights the potentiality of such 3D-2D heterostructures as vital components for fuel cell technologies and/or production of renewable fuels.

**2. Results and discussion**

Key-point in the development of the functional 3D-2D heterostructure CNHs-$MoS_2$ in electrocatalysis is the covalent linkage between the two nanostructured materials. Considering that neither CNHs nor $MoS_2$ contain dangling bonds, activation by incorporation of suitable linkers carrying complementary functional groups, at the conical tips and sidewalls of CNHs as well as at the basal plane of $MoS_2$, is required. In addition, the covalent interactions tightly holding together CNHs and $MoS_2$, compared to the weaker supramolecular or van der Waals ones, significantly perturb their electronic properties within the heterostructure. Furthermore, they also provide structural local matching, between $MoS_2$ and the conical nanotubules forming the aggregated secondary superstructure of CNHs, in the 2D direction. All these, together with the intrinsic high porosity of CNHs, enhance the density of the electrocatalytic active sites within the 3D-2D heterostructure CNHs-$MoS_2$, facilitating protons reduction as it is shown in this work.

The realization of 3D-2D heterostructure CNHs-$MoS_2$, with increased $MoS_2$ loading onto CNHs, is based on a controlled stepwise functionalization route. For this purpose, separately $MoS_2$ and CNHs were pre-modified so that to introduce appropriate organic functional units for precisely connecting the two species via robust covalent bonding. In more detail, starting with the pre-modification of 2D $MoS_2$, exfoliated 1T-$MoS_2$ nanosheets were obtained upon treatment of the bulk material with n-BuLi, followed by hydration and aqueous sonication of the intercalated species.[21] Subsequently, 1T-$MoS_2$ was reacted with in-situ generated diazonium salts[21,22] derived from aniline derivative **1** to yield $MoS_2$-based material **2**. Treatment of the latter with trifluoracetic acid, followed by neutralization of the adduct by triethylamine, produced amino-modified $MoS_2$ nanosheets **3** (**Scheme 1**). The amino-loading of **3** was calculated by the Kaiser test and found to be 100 µmol/g. On a separate functionalization route, the basal plane of exfoliated 1T-$MoS_2$ nanosheets was decorated with carboxylic acid units to yield material **4**, via reaction of 1T-$MoS_2$ with in-situ generated diazonium salts derived from 4-(4-aminophenyl)butyric acid. Then, continuing with the pre-modification of CNHs, light-assisted oxidation with hydrogen peroxide,[23] furnished oxidized CNHs **5**, featuring carboxylic acid moieties at their conical tips. Next, with those properly modified $MoS_2$ nanosheets **3** and CNHs **5** in hand, having complementary functional units for directing their coupling, we proceeded with the conjugation. Condensation reaction between amino-modified $MoS_2$ **3** and oxidized CNHs **5** yielded hybrid **6** (Scheme 1). The amino-loading of **6** was dropped to 15 µmol/g (Kaiser test), verifying the success of the condensation reaction. In order to increase the loading of 1T-MoS$_2$ onto CNHs in the heterostructure, we performed a second functionalization reaction onto the sidewalls of CNHs participating in conjugate **6**. Specifically, reaction of in-situ generated azomethine ylides,[6] derived from the thermal condensation reaction between tert-butyloxycarbonyl (BOC)-protected α-amino acid **7** and formaldehyde, followed by acidic deprotection of the BOC moiety and subsequent neutralization, yielded heterostructure **8**, featuring doubly modified CNHs carrying MoS$_2$ at the tips and fused pyrrolidine rings at the sidewalls being N-alkylated and terminated to amine groups. The amino-loading of **8** was calculated by the Kaiser test and found to be 80 µmol/g. At this point, condensation of those free amine groups with the carboxylic acid units present in modified MoS$_2$ nanosheets **4** resulted on additional incorporation of MoS$_2$, this time at the sidewalls of CNHs, yielding heterostructure **9** (Scheme 1). Again, the drop of amino-loading in **9**, being 15 µmol/g (Kaiser test), guarantees the efficiency of the condensation reaction.

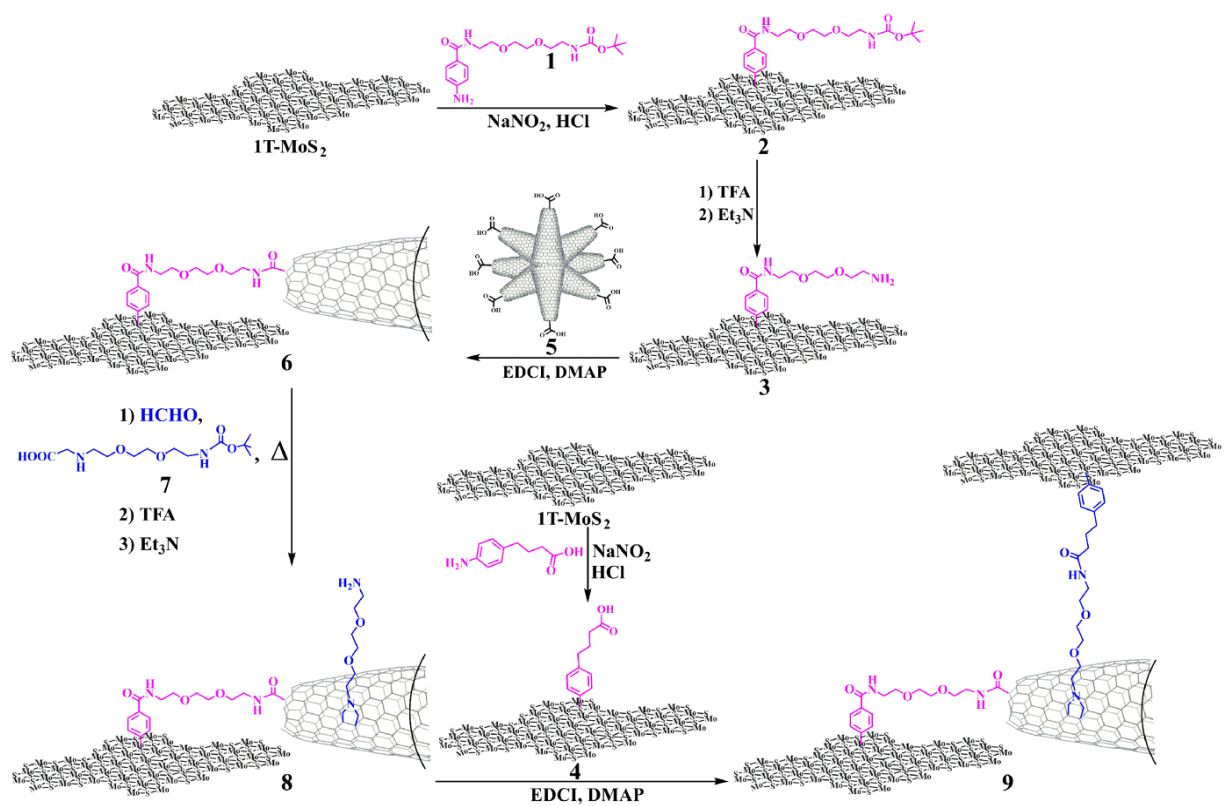

**Scheme 1.** Synthetic illustration of the preparation of CNH-MoS$_2$ heterostructure.

In order to follow every different modification occurred on MoS$_2$ and CNHs, with ultimate aim to verify the successful formation of the 3D-2D heterostructure CNH-MoS$_2$ **9**, diverse spectroscopic, thermal and microscopy imaging techniques were employed. Specifically, the FT-IR spectrum of BOC-modified MoS$_2$ material **2** is controlled by observing the characteristic bands due to carbonyl stretching of benzamide and BOC protecting group at 1645 cm$^{-1}$ and 1691 cm$^{-1}$, respectively, while alkyl C-H vibrational features are discernable in the range 2840–2980 cm$^{-1}$ (Supporting Information, Figure S1). Markedly, the carbonyl BOC band is absent in the FT-IR spectrum of amino-modified MoS$_2$ nanosheets **3**, where particularly vibrations only due to the carbonyl stretching of the benzamide at 1645 cm$^{-1}$ are evident. On the other hand, the FT-IR spectrum of oxidized CNHs **5** is occupied by the characteristic carbonyl vibration mode at 1705 cm$^{-1}$ due to the surface -COOH units (Supporting Information, Figure S2). Heterostructure **9** shows enhanced the band at 1640 cm$^{-1}$, as compared to that observed for **6** (**Figure 1a**), due to the increased carbonyl amides, e.g. derived by condensation and subsequent incorporation of additional MoS$_2$ species at the sidewalls of CNHs.

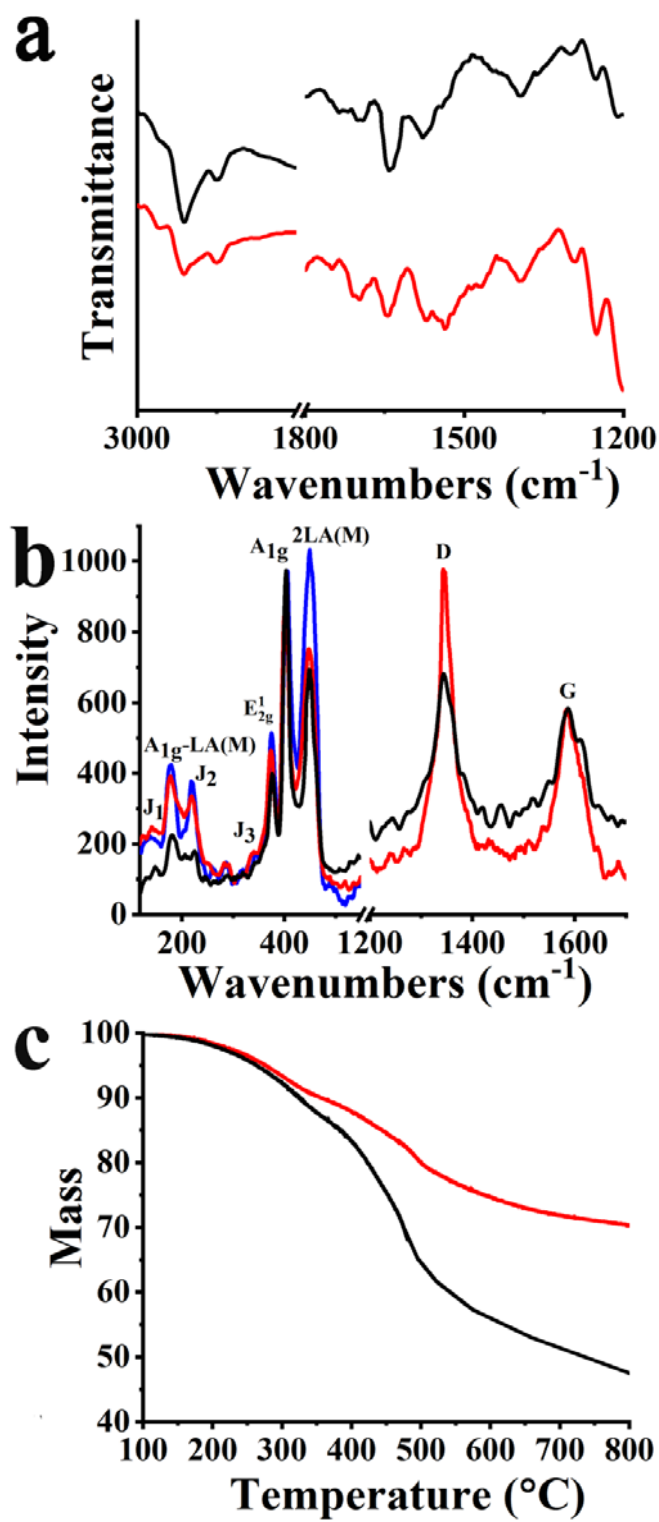

**Figure 1.** (a) ATR-IR spectra for **6** (red) and CNH-MoS$_2$ heterostructure **9** (black). (b) Raman spectra for **6** (red), CNH-MoS$_2$ heterostructure **9** (black) and exfoliated 1T-MoS$_2$ (blue). The MoS$_2$-centered part of the spectrum (120-550 cm$^{-1}$) is recorded upon excitation at 633 nm, while the CNH-based part of the spectrum (1200-1700 cm$^{-1}$) is recorded upon excitation at 514 nm. The spectra shown at the 1200-1700 cm$^{-1}$ area are normalized at the G-band and their intensity is enhanced compared to the MoS$_2$-centered part of the spectrum (120-550 cm$^{-1}$) for clarity. (c) TGA graphs for **6** (red) and CNH-MoS$_2$ heterostructure **9** (black), obtained under nitrogen atmosphere.

Additional insight on the formation of CNHs-MoS$_2$ heterostructure **9** was acquired by Raman spectroscopy. By employing on-resonance excitation (633 nm), with respect to the energy of the A-exciton,[24] and adjusting the laser power to 0.3 mW/cm$^2$ and exposure to 10 seconds to avoid overheating of the samples, we examined all modified MoS$_2$-based materials. Specifically, the spectrum of exfoliated 1T-MoS$_2$ is governed by the A$_{1g}$-LA(M), E$^1_{2g}$, A$_{1g}$ and 2LA(M) Raman active modes at 177, 373, 404 and 450 cm$^{-1}$, respectively. In addition, J$_1$, J$_2$ and J$_3$ bands located at 145, 220 and 340 cm$^{-1}$, respectively, are fingerprint phonon modes of the metallic 1T octahedral phase of MoS$_2$ (Figure 1b). Notably, from those J phonon modes the J$_2$ is the most intense one, and discernible in all modified MoS$_2$-based materials **2**-**4** (Supporting Information, Figure S3), proving the stabilization of the octahedral 1T-MoS$_2$ phase upon aryl diazonium salts functionalization, in agreement with earlier studies.[15,20] Moreover, the intensity ratio 2LA(M)/A$_{1g}$ can be used to assess the basal plane functionalization of 1T-MoS$_2$.[21] Specifically, the 2LA(M)/A$_{1g}$ intensity ratio for materials **2**-**4** appeared decreased to ca. 0.75 due to the basal plane chemical modification as compared to the value of 1.07 for exfoliated 1T-MoS$_2$ (Supporting Information, Figure S3). The particular ratio remained unchanged not only after the removal of the BOC group in material **3**, but also in hybrid **6**, since conjugation of 1T-MoS$_2$ at the tips of oxidized CNHs **5** has no further impact on the basal plane of the MoS$_2$ nanosheets (Figure 1b).

Additionally, in order to determine the functionalization degree in CNHs graphitic network, Raman spectroscopy was performed upon excitation at 514 nm. The Raman spectrum of oxidized CNHs **5** is dominated by the characteristic D- and G- bands at 1340 and 1592 cm$^{-1}$, related to defects and the

sp$^2$ graphitic network, respectively. Most importantly, alterations of the intensity of the D-band are strong indicators of the functionalization degree in CNHs.[25] The D/G intensity ratio was increased to 1.44 for oxidized CNHs **5** (Supporting Information, Figure S4) compared to the value of 1.14 for pristine CNHs due to the generation of defected sites at CNHs tips.[23] This value was found unchanged for hybrid **6** since the conjugation of 1T-MoS$_2$ at the tip of oxidized CNHs **5** does not induce any further alteration on the graphitic network (Supporting Information, Figure S4). On the opposite, heterostructure **8** showed an increase of the I$_D$/I$_G$ ratio, ca. 1.55, due to the second covalent functionalization resulting to new defected sites at the sidewalls of CNHs (Supporting Information, Figure S4). Markedly, the I$_D$/$_G$ ratio remained the same for CNH-MoS$_2$ heterostructure **9**, considering that the interface of additional MoS$_2$ nanosheets occurs via conjugation on the already grafted organic chains present at the sidewalls of CNHs (Figure 1b).

The impact of functionalization on the lattice disorder of MoS$_2$ is further corroborated by performing spatial Raman mapping measurements and accordingly probing the 2LA(M)/A$_{1g}$ ratio. In this context, Raman maps (633 nm) were constructed by acquiring and collecting 120 point-spectra (30 μm x 30 μm area) for 1T-MoS$_2$ and modified MoS$_2$-based materials **2**, **3** and **6** (**Figure 2**) as well as **4** (Supporting Information, Figure S5). The average 2LA(M)/A$_{1g}$ ratio for **2-4** and **6** is by around 30 % lower as compared to that of exfoliated 1T-MoS$_2$ due to symmetry disorder and proving enhancement of the local strain to basal plane modification. Analogously, the structural characteristics of CNHs component in heterostructure **9** were elucidated by Raman mapping assays (514 nm) by collecting 120 point-spectra (30 μm x 30 μm area) and compared to pristine CNHs at **Figure 3**. The corresponding Raman maps for materials **5-8**, for evaluating the defects induced at the lattice as a result of the chemical modification, are shown at the Supporting Information, Figure S6. The average D/G ratio for oxidized CNHs **5**, ca. 1.44, is by around ~21 % higher as compared to that of pristine CNHs, ca. 1.14, resulted from the incorporation of -COOH units at the tips of CNHs. Meanwhile, for **8** and 3D-2D heterostructure **9**, the average D/G ratio, ca. 1.55, is by around ~7 % higher as compared to that of **6**, ca. 1.45, due to disorder of the graphitic network brought by hybridization change from sp$^2$ to sp$^3$ as resulted by the additional grafting of addends and subsequent interface of MoS$_2$ at the sidewalls of CNHs.

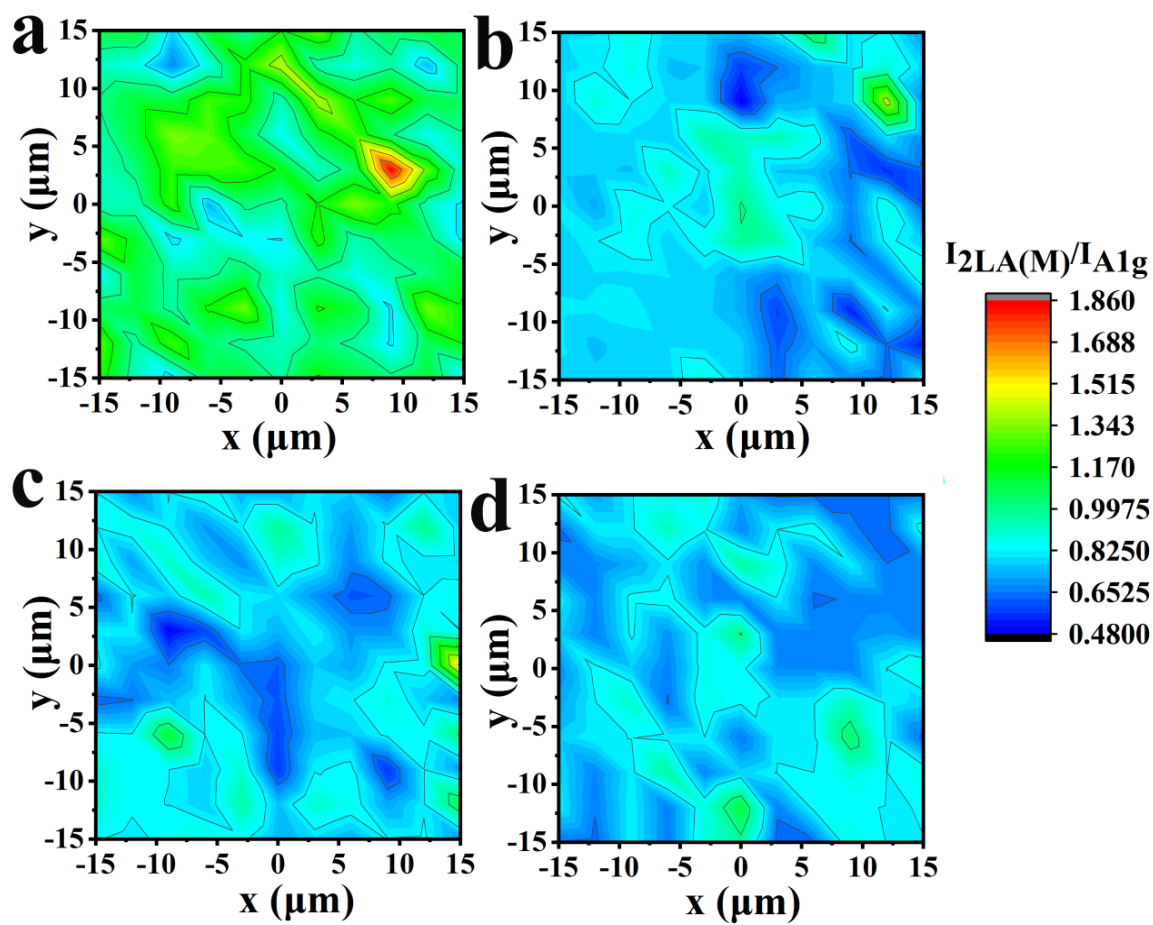

**Figure 2**. Raman maps of (a) exfoliated 1T-MoS$_2$, (b) BOC-modified MoS$_2$ material **2**, (c) amino-modified MoS$_2$ material **3**, and (d) material **6**, recorded upon excitation at 633 nm.

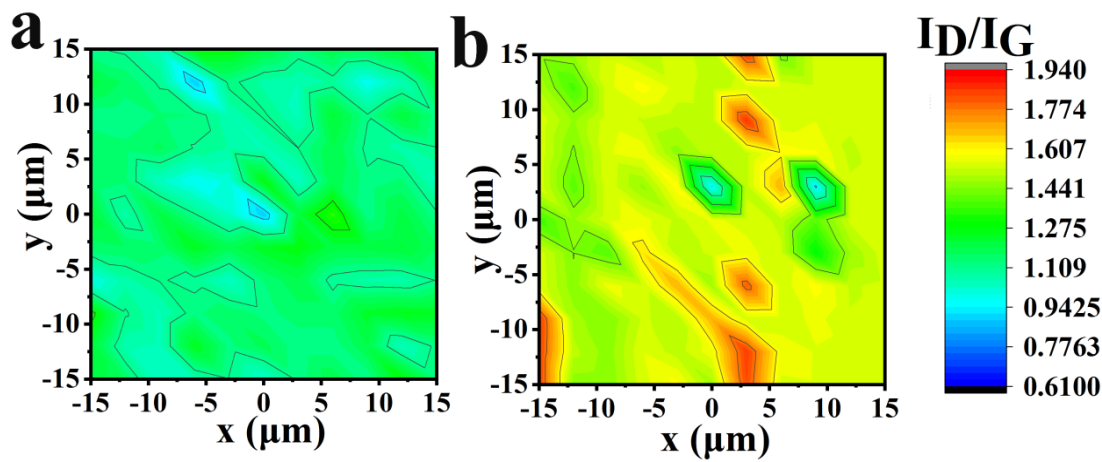

**Figure 3**. Raman maps of (a) pristine CNHs, and (b) 3D-2D heterostructure **9** recorded upon excitation at 514 nm.

XPS was also employed to get additional information about the chemistry performed and identify the atomic species corresponding to the addends that link CNHs and MoS$_2$ in heterostructure **9**. In more detail, **Figure 4a** shows a survey XPS spectrum for the CNH-MoS$_2$ material **9**, where the existence of C, Mo, S and O is confirmed. High-resolution spectra of these elements have been acquired and also depicted in this figure. The S 2p core-level spectrum is shown in Figure 4b, where two strong peaks at 162.0 and 163.2 eV are observed. These peaks correspond to S 2p$_{3/2}$ and S 2p$_{1/2}$ binding energies for S$^{2-}$. Figure 4c displays the XPS spectrum in the region of the S 2s and Mo 3d orbitals. The deconvolution of the different peaks discerned in this spectrum is as follows: the peak at 226.3 eV corresponds to S 2s and is assigned to MoS$_2$ nanosheets; the second peak at 229.0 eV corresponds to Mo$^{4+}$ 3d$_{5/2}$; the third peak at ~232.3 eV is the contribution of two different configurations of molybdenum, those of Mo$^{4+}$ 3d$_{3/2}$ and Mo$^{6+}$ 3d$_{3/2}$; the last peak at 235.6 eV is due to Mo$^{6+}$ 3d$_{5/2}$. Thus, two different molybdenum contributions can be detected, namely, Mo$^{4+}$ representing around 90% of the total amount and assigned to MoS$_2$ (as expected in the 1T phase),[26] and Mo$^{6+}$ of around 10% corresponding to MoO$_3$. This small amount of oxide suggests that during the different chemical transformations and/or treatments developed on these materials, some oxygen is incorporated in the outermost layer of MoS$_2$, hindering further diffusion to the flakes and protecting them from undesired deeper oxidation. It is important to point out that the quantitative analysis of the S/Mo$^{4+}$ (equal to 1.99) in the Mo 3d S 2s region confirms the good stoichiometry for MoS$_2$ within heterostructure **9**. Figure 4d displays the C 1s part of the XPS spectrum, with the peaks at 284.5, ~284.9, 286.2 and 288.1 eV corresponding to C=C and C-C, COOH and C=O, respectively.[27] Finally, in Figure 4e, the N 1s contribution at 399.9 eV has been observed, confirming the presence of nitrogen in the linkers that join the two nanostructures in CNH-MoS$_2$ material **9**.

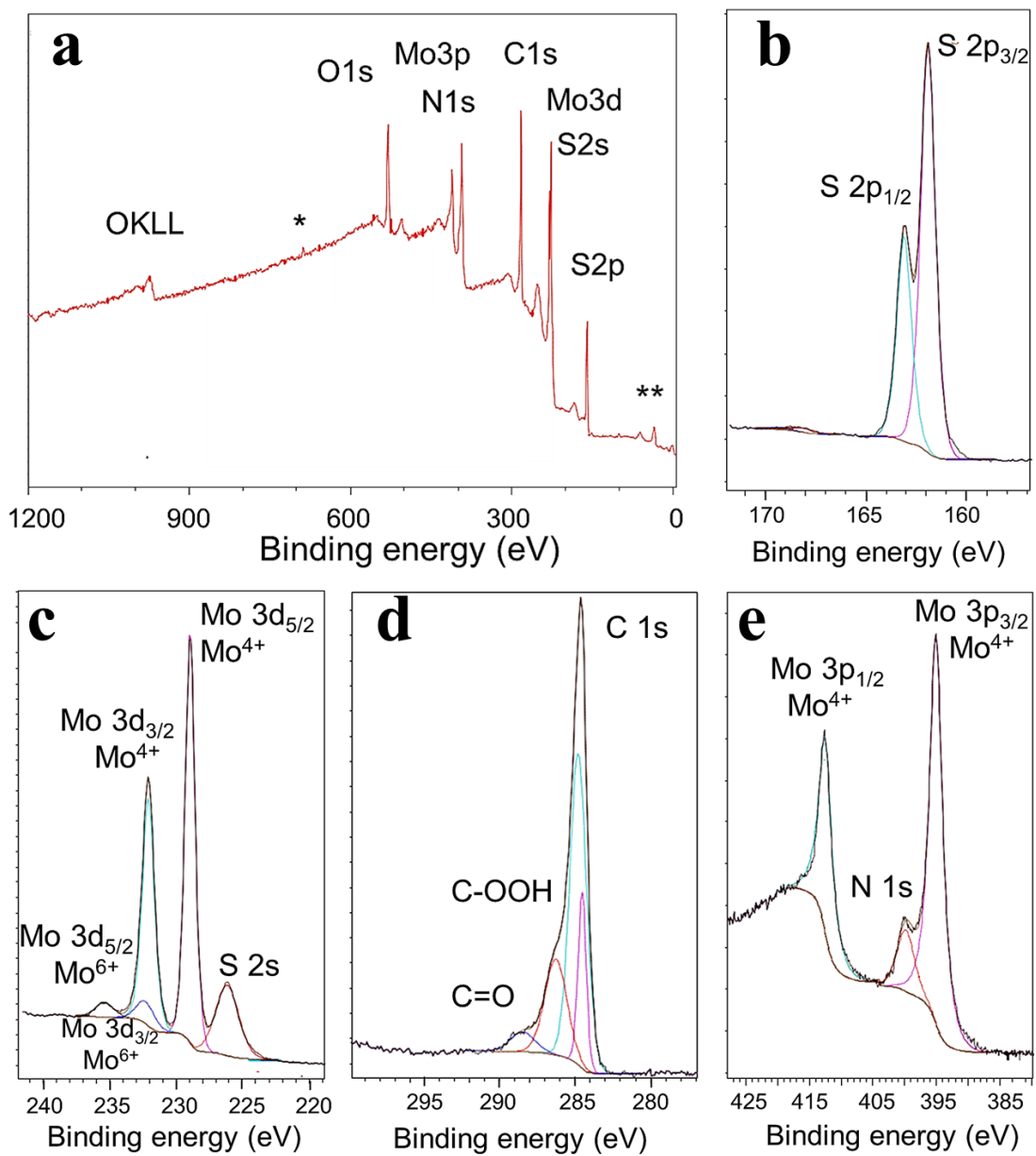

**Figure 4.** (a) XPS survey spectrum of CNH-MoS$_2$ heterostructure **9**, where S 2p, S 2s, Mo 3d, C 1s, N 1s, Mo 3p and O 1s are observed. Impurities < 1% corresponding to F and Si are shown with * and **,

respectively. (b-e) High-resolution XPS spectra of CNH-MoS$_2$ heterostructure **9**, recorded on the same area, and corresponding to S 2p, S 2s-Mo 3d, C 1s and Mo 3p-N 1s, respectively.

Structural and chemical analyses, at the subnanometer scale, have been developed via different (scanning) transmission electron microscopy ((S)TEM) techniques, see **Figure 5**. The combination of different TEM techniques allow to perform detailed analysis of these nanostructures.[12,14,16,23,28] Figure 5a,b display two low magnification micrographs corresponding to MoS$_2$ nanosheets and to the heterogeneous system of CNHs and nanosheets in **9**. These images show how both nanostructures are organized. It should be mentioned that TEM observations confirm the highest abundance of MoS$_2$ with respect to CNHs. These nanomaterials can be clearly observed in the HRTEM images of Figure 5c-e; see also Supporting Information, Figure S7 for images corresponding to hybrid material **8** as well as exfoliated MoS$_2$. As above mentioned (see Scheme 1), the MoS$_2$ nanosheets are seen both anchored at the tips to CNHs as well as incorporated on their sidewalls. Figure 5f corresponds to a HAADF-STEM image of one of these nanostructures confirming what it was observed by HRTEM (Figure 5c-5e). In the white rectangular region marked in that image of Figure 5f, a STEM-EELS spectrum-image was acquired. Figure 5g shows a combination of chemical maps of two of the elements detected in that area, C and S, which correspond to CNHs and to MoS$_2$, respectively.

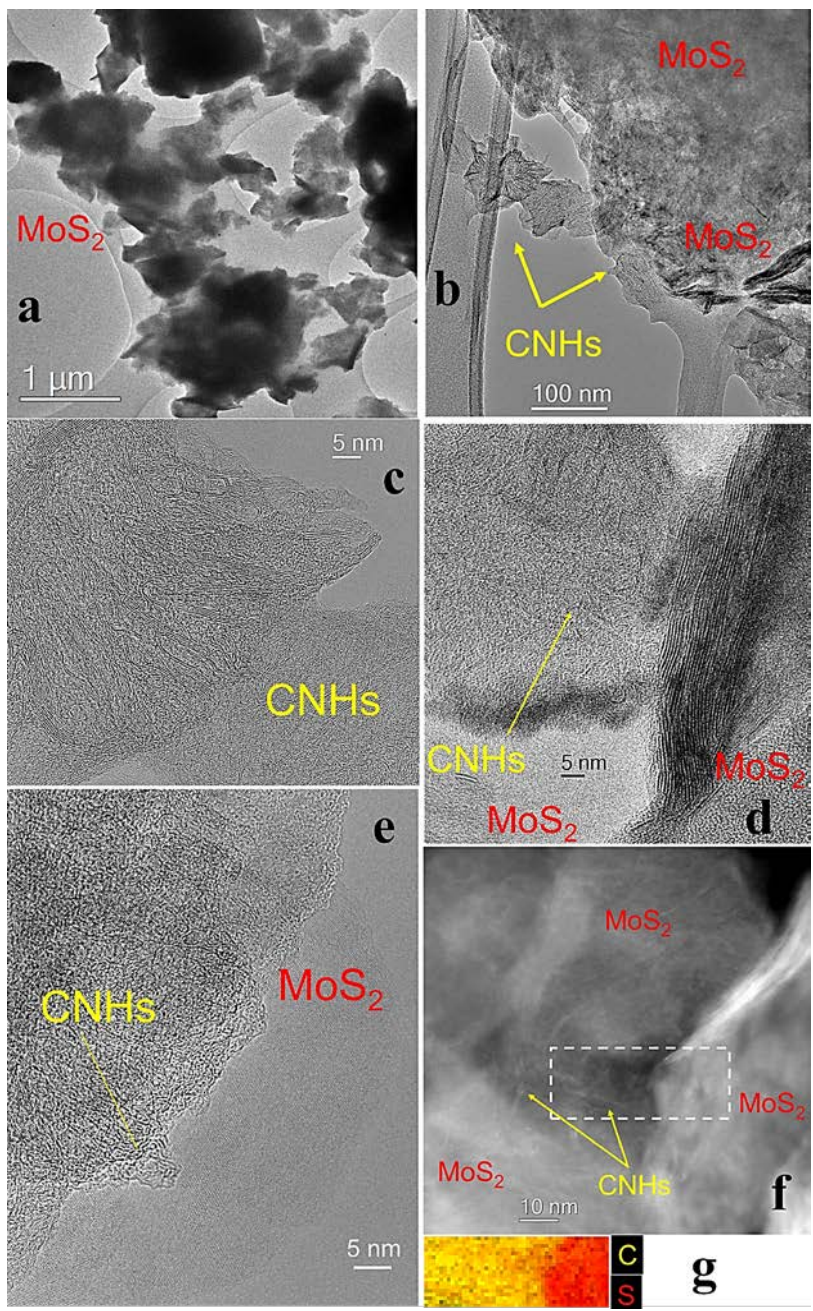

**Figure 5**. (a, b) Low-magnification TEM micrographs of CNH-MoS$_2$ heterostructure **9**, where (a) MoS$_2$ nanosheets and (b) hybrid nanostructures (CNHs and MoS$_2$ flakes) can be observed respectively. (c-e) HRTEM micrographs displaying the distribution of the two constituents of these heterostructures, namely CNHs and the MoS$_2$ nanosheets. (f) HAADF-STEM image showing one of these CNH-MoS$_2$ heterostructures. A STEM-EELS spectrum-image has been recorded in the rectangular dashed white area highlighted in Fig. 5f. (g) The combined colour chemical map showing C (in yellow; from the CNHs) and S (in red; corresponding to the MoS$_2$ nanosheets) distribution.

For collecting information about the thermal stability of heterostructure **9** and all intermediate modification steps for CNHs and MoS$_2$, as well as estimating the amount of organic units covalently anchored onto the 3D and 2D nanostructures, thermogravimetric analysis (TGA) under N$_2$ atmosphere was performed. In this context, exfoliated 1T-MoS$_2$ presents a 4% mass loss up to 500 °C due to the presence of defects related to the preparation method,[21] while BOC-modified MoS$_2$ material **2** shows a higher mass loss of 16% up to same temperature region due to the thermal decomposition of the incorporated organic species (Supporting Information, Figure S8). This percentage is lower for the amino-modified MoS$_2$ material **3**, ca. 7%, as expected, due to the absence of the BOC protecting group that contributes to the system with higher molecular weight. Based on that, the loading in the amino-modified MoS$_2$ material **3** was calculated to be one functional group for every 15 MoS$_2$ units. Meanwhile, the thermograph of material **4** exhibits a 10 % mass loss up to 500 °C, also corresponds to 1 functional unit per every 15 MoS$_2$ units (Supporting Information, Figure S8). At the same time, pristine CNHs present an almost stable thermal profile under nitrogen atmosphere up to 800 °C, while oxidized CNHs **5** lose 9% of mass up to 500 °C, due to the thermal decomposition of the carboxylic acid groups (Supporting Information, Figure S9). The thermal decomposition at higher temperature is ascribed to the gradual destruction of the graphitic skeleton, occurring at sp$^3$ hybridized sites, where functionalization took place. In addition, the thermograph of material **6** reveals a higher mass loss of 27 % up to 500 °C, showing the successful hybridization of oxidized CNHs **5** with modified MoS$_2$ material **3** (Figure 1c). The subsequent second modification of CNHs at the sidewalls via the addition of in-situ generated azomethine ylides is screened with a mass loss of 25% up to 500 °C for heterostructure **8** (Supporting Information, Figure S8). Lastly, the increased mass loss of 46 % in the thermograph of heterostructure **9** compared to that of **6** (Figure 1c) advocates the successful addition of MoS$_2$ nanosheets onto the sidewalls of CNHs.

In order to validate the aspirations of our aim, we constructed pseudoelectrodes by drop-casting dispersions of the 3D-2D heterostructure **9** onto a glassy-carbon electrode and assessed their activity for protons reduction in a proof-of-concept application. Markedly, comparing the electrocatalytic behavior of **9** with that of **6** toward HER, the beneficial role of higher loading of MoS$_2$ nanosheets, covalently anchored at the sidewalls of the doubly-modified CNHs, was revealed. In more detail, performing linear sweep voltammetry (LSV) assays in aqueous 0.5 M H$_2$SO$_4$ electrolyte, an onset overpotential at 0.029 V vs RHE was recorded for CNH-MoS$_2$ heterostructure **9** (**Figure 6a**), which is by 249 mV lower than the one recorded for **6**, ca. -0.220 V vs RHE. Notably, the excess energy above the needed thermodynamic potential to drive protons reduction to molecular hydrogen, i.e. onset overpotential, for **9** is same with the one noted for Pt/C employed as reference material, highlighting the advantageous nature of the newly prepared heterostructure. Furthermore, considering that the functional current density required for electrochemical water splitting and sufficient hydrogen production is -10 mA/cm$^2$, while misinterpretations from inherent electroactivity can be avoided at that value, it is reasonable to use the potential required to reach the aforementioned current density as point of reference for protons reduction by **9** and compare it with that of **6**. Specifically, while for **6** an overpotential of -0.340 V vs RHE was registered at -10 mA/cm$^2$ current density, the 3D-2D heterostructure **9** operates the HER at -0.028 V vs RHE, ca. 312 mV lower than that of **6**, and close to the overpotential registered for the reference Pt/C at -0.009 V vs RHE (Figure 6a). The significantly lower overpotential value for driving protons reduction to molecular hydrogen with heterostructure **9** over **6** is directly related to the presence of increased electrocatalytic centers and clearly mirrors the higher loading of MoS$_2$ nanosheets at both the tips and sidewalls of CNHs. The superior electrocatalytic activity of heterostructure **9** toward protons reduction is ascribed to the electronic communication between the two covalently linked species within the 3D-2D nanoconjugate and the plethora of MoS$_2$-based active centers incorporated onto the extended π-electronic conductive network of CNHs. The latter is further corroborated by considering the significantly inferior HER electrocatalytic activity of heterostructures prepared by direct growth of MoS$_2$ onto graphene[12,13,29-32] or via electrostatic interactions between MoS$_2$ and graphene.[33] From the LSVs of MoS$_2$-based material **3**, exfoliated 1T-MoS$_2$ and oxidized CNHs **5** (Supporting Information, Figure S10a), higher overpotential values compared to that of **9** for HER are noted. The slightly weakened electrocatalytic activity of **3** versus exfoliated 1T-MoS$_2$ is rationalized by considering the removal/neutralization of negative charges from the surface of 1T-MoS$_2$ due to the functionalization and formation of covalent S–C bonds.[15]

Insight on the reaction kinetics and the exchange current for the electrocatalytic protons reduction with heterostructure **9** was obtained by the Tafel slope extracted from the corresponding LSV curve. Briefly, in electrochemical kinetics, the Tafel slope relates the rate of HER to the overpotential and can give useful information about how quickly the current increases with the applied overpotential. With the above in mind, heterostructure **9** possesses a low Tafel slope value of 71 mV/dec (Figure 6b), which manifests that hydrogen production is rate-limited by the electrochemical desorption of adsorbed hydrogen atoms onto the modified pseudoelectrode. On the contrary, **6** exhibits a high Tafel slope value of 161 mV/dec (Figure 6b), revealing different rate-limiting step for HER, namely adsorption of protons onto the pseudoelectrode surface via a reduction process. The Tafel slope values for MoS$_2$-based material **3**, exfoliated 1T-MoS$_2$ and oxidized CNHs **5** (Supporting Information, Figure S10b) are also high, ca. 150, 153 and 274 mV/dec, respectively.

Additional insight on HER kinetics is obtained by electrochemical impedance spectroscopy (EIS) assays. The EIS measurements were conducted at a potential where significant HER current was recorded, corresponding to -2 mA/cm$^2$, while EIS data were fitted to Randles circuit. Briefly, the low frequency region in EIS was employed to determine the charge-transfer resistance (R$_{ct}$) at the interface with the electrolyte. Thus, Nyquist plot showed the smallest frequency semicircle for heterostructure **9**, attributed to small R$_{ct}$ of 34.5 Ω (Figure 6c) as a result of high conductance derived by the covalently incorporated CNHs in the 3D-2D heterostructure. On the contrary, **6** showed R$_{ct}$ of 79.0 Ω, higher than the value calculated for exfoliated 1T-MoS$_2$ at 57.0 Ω (Supporting Information, Figure S10c). The higher conductance of **6** compared to exfoliated 1T-MoS$_2$, as deduced from EIS and the calculated R$_{ct}$ values, is rationalized by considering the removal/neutralization of the negative charges in MoS$_2$ due to the aryl diazonium salts functionalization leading to lower conductance. In fact, this is validated by examining the R$_{ct}$ of modified MoS$_2$ nanosheets **3**, calculated to be 72.0 Ω, similar to that of **6** and higher than that of exfoliated 1T-MoS$_2$ due to lower conductance resulted from the quenching of the negative surface charge upon chemical modification.[15] Overall, the results from EIS as presented by the calculated R$_{ct}$ agree with the HER kinetics and Tafel slope values for the efficient performance of heterostructure **9**.

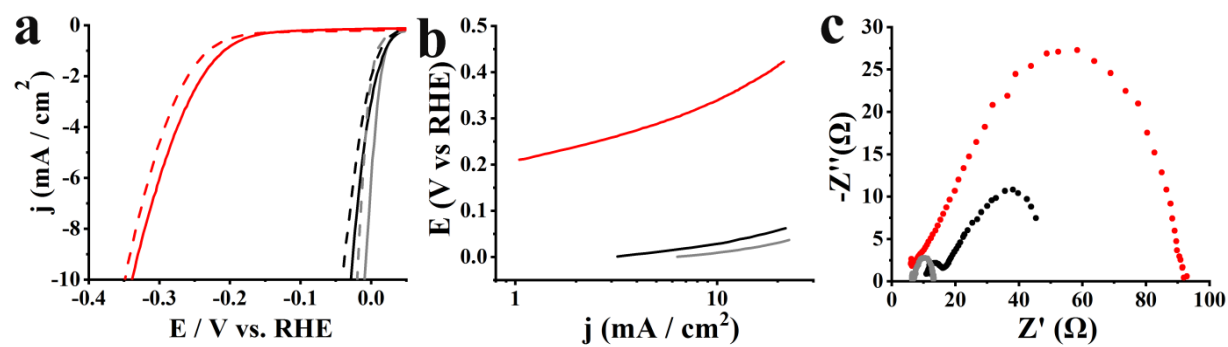

**Figure 6.** (a) LSVs for HER before (solid lines) and after 10,000 cycles (dashed lines), (b) Tafel slopes, and (c) Nyquist plots, for materials **6** (red), **9** (black) and Pt/C (grey). The LSVs obtained at 1,600 rpm rotation speed and 5 mV/s scan rate in aqueous 0.5 M H$_2$SO$_4$.

The stability of **9** was assessed, after performing 10,000 ongoing electrocatalytic cycles. Specifically, heterostructure **9** proved to be extremely stable after 10,000 cycles by exhibiting a negatively shifted LSV curve of 11 mV (Figure 6a). The same was noticed for Pt/C employed as reference. On the contrary, **6** as well as exfoliated 1T-MoS$_2$, modified MoS$_2$ nanosheets **3** and oxidized CNHs **5** showed higher overpotentials of 40-90 mV after continues cycling (Supporting Information, Figure S10a). Table 1 summarizes the electrochemical HER data before and after 10,000 cycles for heterostructure **9** and compares with those of **6** as well as exfoliated 1T-MoS$_2$ and oxidized CNHs **5**.

**Table 1.** Electrocatalytic HER parameters for heterostructure **9** in comparison with materials **3**, **6**, oxidized CNHs **5**, exfoliated 1T-MoS$_2$ and Pt/C.

| Electrocatalyst | Onset potential (V vs RHE) | Potential (V vs RHE) at -10 mA/cm$^2$ | Tafel slope (mV/dec) | R$_{ct}$ (Ω) |
|---|---|---|---|---|
| Heterostructure 9 | 0.029 | -0.028 | 71 | 34.5 |
| Heterostructure 9[a] | 0.02 | -0.039 | 67 | - |
| Pt/C | 0.029 | -0.009 | 35 | 6.1 |
| Pt/C[a] | 0.011 | -0.020 | 35 | - |
| 6 | -0.22 | -0.34 | 161 | 79 |
| 6[a] | -0.25 | -0.43 | 209 | - |
| Exfoliated 1T-MoS$_2$ | -0.16 | -0.28 | 153 | 57.0 |
| Exfoliated 1T-MoS$_2$[a] | -0.21 | -0.32 | 150 | - |
| 3 | -0.18 | -0.31 | 150 | 72 |
| 3[a] | -0.25 | -0.40 | 165 | - |
| oxidized CNHs 5 | -0.25 | -0.43 | 274 | 197.0 |
| oxidized CNHs 5[a] | -0.29 | -0.48 | 240 | - |

Finally, in order to further assess the beneficial role of covalently linking the species composing the CNH-MoS$_2$ heterostructure, we evaluated the electrocatalytic HER activity of pseudoelectrodes fabricated by physical mixing materials 3 and 5 as well as 4 and 6, abbreviated as mixtures A and B respectively, and compare them with materials 6 and 9, respectively. As expected from the LSVs of the mixed materials (Supporting Information, Figure S11a), extremely higher onset potential values for HER, ca. -0.41 and -0.19 V vs RHE and potential values at -10 mA/cm$^2$ ca. -0.66 and -0.35 V vs RHE compared to that of 6 and 9 ca. -0.22 and 0.029 V vs RHE and ca. -0.34 and -0.028 V vs RHE, respectively, are noted. The poor hydrogen production is also mirrored by the extremely high Tafel slope values i.e. 415 and 211 for mixtures A and B, respectively, versus 161 and 71 for 6 and 9, respectively, highlighting the sluggish current flow within the species (Supporting Information, Figure S11b) due to inadequate and weak interactions. Lastly, the stability of the mixtures was assessed, after performing 10,000 ongoing electrocatalytic cycles. Unsurprisingly, mixtures A and B showed higher onset overpotentials by ca. 272 and 520 mV and potentials registered at -10 mA/cm$^2$ ca. 280 and 681 mV as compared to the values registered for 6 and 9, respectively, after continues cycling (Supporting Information, Figure S11a) due to the absence of strong interactions holding tight the species forming the heterostructures. Overall, the above findings underline the importance of the development of heterostructures, wherein components are firmly bound together, i.e. via covalent linkage, in order to promote charge transfer and current flow but also to ensure stability; all of them being crucial factors for electrocatalysis.

## 3. Conclusions

The combination of nanomaterials with interesting characteristics in the form of advanced heterostructures is a potent approach to boost their properties and overtake limitations that individual components might face. In this work, we demonstrated a novel functionalization approach towards the preparation of the 3D-2D CNH-MoS$_2$ heterostructure, relying on the modification of the conical tips and sidewalls of CNHs to covalently link with chemically transformed MoS$_2$ nanosheets. We carefully followed every modification step and verified the success of performed reactions by FT-IR and Raman spectroscopy, XPS, TGA, and electron microscopy imaging and spectroscopy means. The plethora of catalytic active sites due to the MoS$_2$ incorporation in the 3D modified porous CNH superstructure and the covalent linkage between CNH and MoS$_2$ improving charge-transfer at the heterointerfaces enhanced the electrocatalytic properties of the CNH-MoS$_2$ heterostructure, which showed excellent activity toward protons reduction, same as the one noted for commercial Pt/C. Our results are fully supported by data acquired from linear sweep voltammetry and electrochemical impedance spectroscopy, highlighting minute overpotential, low Tafel slope and small charge-transfer resistance for electrocatalyzing the evolution of hydrogen. Furthermore, the stability of the CNH-MoS$_2$ heterostructure was validated after performing 10,000 ongoing electrocatalytic cycles. All in all, this advanced functionalization approach towards the preparation of the novel 3D-2D heterostructure may open new routes for the realization and exploration of functional heterostructures.

## 4. Experimental Section/Methods

**General.** Chemicals, reagents, and solvents were purchased from Sigma-Aldrich and used without further purification. For the light-assisted oxidation reaction of CNHs the light source used was a 500 W halogen lamp, which was positioned 20 cm away from the reactor. Infrared (IR) spectra were obtained on a Fourier Transform IR spectrometer (Equinox 55 from Bruker Optics) equipped with a single reflection diamond ATR accessory (DuraSamp1IR II by SensIR Technologies). Raman measurements were recorded with a Renishaw confocal spectrometer at 514 and 633 nm. The data were obtained and analysed with Renishaw Wire and Origin software. Thermogravimetric analysis was acquired using a TGA Q500 V20.2 Build 27 instrument by TA in a nitrogen (purity >99.999%) inert atmosphere. HRTEM studies have been carried out in an image-corrected corrected FEI Titan-Cube 60-300 working at 80 kV. Scanning transmission electron microscopy (STEM) and electron-energy loss spectroscopy (EELS) analyses have been developed in a probe-corrected FEI Titan-Low-Base 60-300 operating at 80 kV (equipped with a X-FEG® gun and Cs-probe corrector (CESCOR from CEOS GmbH)). EELS studies have been performed using the spectrum-image mode.[34, 35] The different CNH-MoS$_2$ powders have been dispersed in ethanol and the suspensions have been ultrasonicated and dropped onto copper carbon holey grids. XPS data were acquired using a Kratos Axis Supra spectrometer equipped with a monochromated Al Kα X-ray source using an analyzer pass energy of 160 eV for survey spectra and 20 eV for the core level spectra. Spectra were recorded by setting the instrument to the hybrid lens mode and the slot mode providing approximately a 700 x 300 μm$^2$ analysis area using charge neutralization. Regions have been calibrated using the reference value BE(C 1s sp2) = 284.5 eV.

All XPS spectra were analyzed using CASA XPS software. The XPS peaks were fitted to GL(70) Voigt lineshape (a combination of 70% Gaussian and 30% Lorentzian character), after performing a Shirley background subtraction. All electrochemical measurements were carried out using an Autolab PGSTAT128N potentiostat/galvanostat and were carried out at room temperature in a standard three-compartment electrochemical cell by using a graphite rod as a counter-electrode, an RDE with glassy carbon disk (geometric surface area: 0.0196 cm$^2$) as a working electrode, and Hg/HgSO$_4$ (0.5 M K$_2$SO$_4$) as reference electrode. LSV measurements for HER were carried out at room temperature in N$_2$-saturated aqueous 0.5 M H$_2$SO$_4$. The catalyst ink was prepared by dispersing 4.0 mg of the catalytic powder in a 1 mL mixture of deionized water, isopropanol, and 5% Nafion (v/v/v=4:1:0.02) and sonicated for 30 min prior use. Before casting the electrocatalytic ink on the electrode's surface, the working electrode was polished with 6, 3 and 1 mm diamond pastes, rinsed with deionized water, and sonicated in double-distilled water. Afterwards, 8.5 µL aliquots of the electrocatalyst were casted on the electrode surface and were left to dry at room temperature. Electrochemical impedance spectroscopy (EIS) measurements were conducted from 10$^5$ to 10$^{-1}$ Hz with an AC amplitude of 0.01 V.

**Exfoliated 1T-MoS$_2$.** In a round-bottom flask, bulk MoS$_2$ (1.5 gr) was treated with n-butyllithium (20 mL, 2.5 M in hexane) and allowed to stir under an inert atmosphere for 48 h at 70 °C. The reaction was then quenched with the slow addition of ~100 mL of distilled water. After hydrogen evolution ceased, deionized water (150 mL) was added and the dispersion was bath sonicated for 1 hour, transferred to a beaker and left overnight to settle. Afterwards, the supernatant was collected in order to avoid non-exfoliated MoS$_2$. The dispersion was filtrated through PTFE membrane filter (0.2 µm) and extensively washed with deionized water and hexane. Next, the filter cake was dispersed in methanol and centrifuged at 4000 rpm for 5 minutes (5 times) with methanol to obtain 300 mg of 1T-MoS$_2$ as powder.

**MoS$_2$-based material 2.** In a round-bottom flask, exfoliated 1T-MoS$_2$ (160 mg) was dissolved in deionized H$_2$O (160 mL) and sonicated for 30 min. In a separate round-bottom flask, aniline derivative **1 (**1.0 gr, 2.7 mmol) was dissolved in a mixture of deionized H$_2$O (32 mL) and concentrated HCl (0.4 mL). Then, NaNO$_2$ (789 mg, 11.4 mmol) was slowly added in the solution under vigorous stirring at 0 °C and 20 % HCl (2.4 mL). The reaction was left for 45 minutes to stir. Afterwards, the solution was added in the 1T-MoS$_2$ dispersion at 0 °C under stirring and a black precipitate was formed. The reaction mixture was left to stir for 2 h at 0 °C and then for another 4 h at ambient. The resulting suspension was filtrated through PTFE membrane filter (pore size 0.2 µm) and washed with deionized H$_2$O, methanol and acetone to yield 150 mg of **2** as powder.

**MoS$_2$-based material 3.** Material **2** (150 mg) was dispersed in dichloromethane (100 mL) and treated with trifluoroacetic acid (5 mL), while the dispersion was left to stir overnight. After filtration in a PTFE membrane filter (0.2 µm) and washing with deionized water, acetone and methanol, the filter cake was dispersed in methanol (50 mL). Then, triethylamine (10 mL) was added and the dispersion was left to stir for 15 minutes. Finally, the dispersion was filtered with PTFE membrane filter (0.2 µm) and washed with dimethylformamide, acetone and methanol. The solid residue was collected to obtain 150 mg of **3** as powder.

**MoS$_2$-based material 4.** In a round bottom flask, 1T-MoS$_2$ (120 mg) was dissolved in deionized H$_2$O (120 mL) and the suspension was sonicated for 30 min. In a separate round-bottom flask, commercially available 4-(4-aminophenyl)butyric acid (350 mg, 2 mmol), was dissolved in a mixture of deionized H$_2$O (25 mL) and concentrated HCl (0.3 mL). Then, NaNO$_2$ (135 mg, 1.95 mmol) was slowly added in the solution under vigorous stirring at 0 °C and 20 % HCl (1.9 mL) was added and the reaction was left under stirring for 45 minutes at 0 °C. Afterwards, the solution was added in the 1T-MoS$_2$ dispersion at 0 °C under stirring and a black precipitate was formed. The reaction was left to stir for 2 h at 0 °C and then for another 4 h at ambient. The resulting suspension was filtrated through PTFE membrane filter (pore size 0.2 µm) and washed with deionized water, methanol and acetone to obtain 120 mg of **4** as powder.

**Oxidized CNHs 5.** In a round-bottom flask, pristine CNHs (30 mg) were treated with H$_2$O$_2$ (30 %, 50 mL) under light irradiation at 120 °C for 3 hours. Then, the dispersion was filtrated through PTFE membrane filter (pore size 0.2 µm) and the solid residue was extensively washed with deionized water and methanol to obtain 30 mg of **5** as powder.

**CNH-based material 6.** In a two-neck round bottom flask, material **5** (2 mg) was added in dry dichloromethane (15 mL) and the suspension was bath sonicated for 30 min under $N_2$ atmosphere. Afterwards, 1-ethyl-3-(3-dimethylaminopropyl)carbodiimide (5.13 mg, 0.027 mmol) was added at 0 °C and the resulting suspension was left to stir for 1 h. Then, 1T-$MoS_2$ (14.5 mg), 4-dimethylaminopyridine (4.7 mg, 0.0382 mmol) and N,N-diisopropylethylamine (20 µL) were added and the dispersion was left to stir for five days under $N_2$ atmosphere in R.T. Finally, the resulting suspension was filtrated through PTFE membrane filter (0.2 µm) and washed with dichloromethane, methanol and acetone. The solid residue was collected to obtain 16 mg of **6** as powder.

**Material 8.** Material **6** (15 mg) was dissolved in DMF (15 mL) and an excess of α-amino acid **7** (17 mg, 0.06 mmol) and formaldehyde (8 mg, 0.27 mmol) were added. The reaction mixture was heated at 120 °C for 3 days. Every 24 hours an excess of **7** (17 mg, 0.06 mmol) and formaldehyde (8 mg, 0.27 mmol) was added. Afterwards, the mixture was filtrated through PTFE membrane filter (0.2 µm), washed with DMF and dichloromethane to obtain 15 mg as powder. Then, all material was added in dichloromethane (20 mL) and treated with trifluoroacetic acid (5 mL), under stirring overnight. After filtration in a PTFE membrane filter (0.2 µm) and washing with deionized water, acetone and methanol, the filter cake was dispersed in methanol (50 mL). Then, triethylamine (10 mL) was added and the mixture was left to stir for 15 minutes. Finally, the suspension was filtered over PTFE membrane filter (0.2 µm) and washed with DMF, acetone and methanol to obtain 15 mg of **8** as powder.

**3D-2D heterostructure 9.** In a two-neck round bottom flask, material **8** (9 mg) was added in dry dichloromethane (15 mL) and sonicated for 30 min under $N_2$ atmosphere. Afterwards, EDCI (3.22 mg, 0.0168 mmol) was added at 0 °C and the suspension was left to stir for 1 h. Then, $MoS_2$-based material **4** (3.1 mg), DMAP (4.7 mg, 0.024 mmol) and N,N-diisopropylethylamine (20 µL) were added and the dispersion was left to stir for five days at room temperature. Finally, the resulting suspension was filtrated through PTFE membrane filter (0.2 μm) and washed with dichloromethane, methanol and acetone to obtain 12 mg of **9** as powder.

**Supporting Information**

Supporting Information is available from the Wiley Online Library or from the author.

**Acknowledgements**

This research is co-financed by Greece and the European Union (European Social Fund) through the Operational Programme "Human Resources Development, Education and Lifelong Learning" in the context of the project "Reinforcement of Postdoctoral Researchers – 2nd Cycle" (MIS 5033021), implemented by the State Scholarships Foundation (IKY). R.A. gratefully acknowledges the support from the Spanish MICINN (PID2019-104739GB-100/AEI/10.13039/501100011033), Government of Aragon (project DGA E13-17R (FEDER, EU)) and from the European Union H2020 programs "ESTEEM3" (Grant number 823717) and Graphene Flagship (881603). Partial financial support from the project "National Infrastructure in Nanotechnology, Advanced Materials and Micro-/Nanoelectronics" (MIS 5002772), which is implemented under the Action "Reinforcement of the Research and Innovation Infrastructures", funded by the Operational Program "Competitiveness, Entrepreneurship and Innovation" (NSRF 2014-2020), Ministry of Development and Investments, and co-financed by Greece and the European Union (European Regional Development Fund) is also acknowledged. TEM and XPS measurements were performed at the Laboratorio de Microscopias Avanzadas (LMA) - Universidad de Zaragoza (Spain). We thank G. Antorrena (LMA) for his help with the XPS data acquisition.

The synthesis of 3D-2D CNH-MoS$_2$ heterostructures was accomplished. The increased MoS$_2$ loading, covalently bound at the tips and sidewalls of CNHs and the covalent linkage between CNH and MoS$_2$ facilitate electrocatalytic activity for protons reduction at heterointerfaces same to that of commercial Pt/C.

*A. Kagkoura,[1] R. Arenal,[2,3,4] N. Tagmatarchis[1*]*

**Controlled chemical functionalization toward 3D-2D carbon nanohorn-MoS$_2$ heterostructures with enhanced electrocatalytic activity for protons reduction**

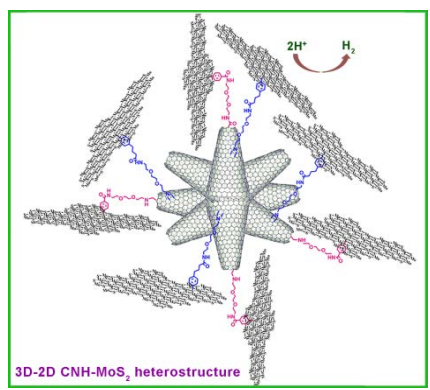

Supporting Information

# Controlled chemical functionalization toward 3D-2D carbon nanohorn-MoS$_2$ heterostructures with enhanced electrocatalytic activity for protons reduction

*Antonia Kagkoura,[1] Raul Arenal,[2,3,4] Nikos Tagmatarchis[1*]*

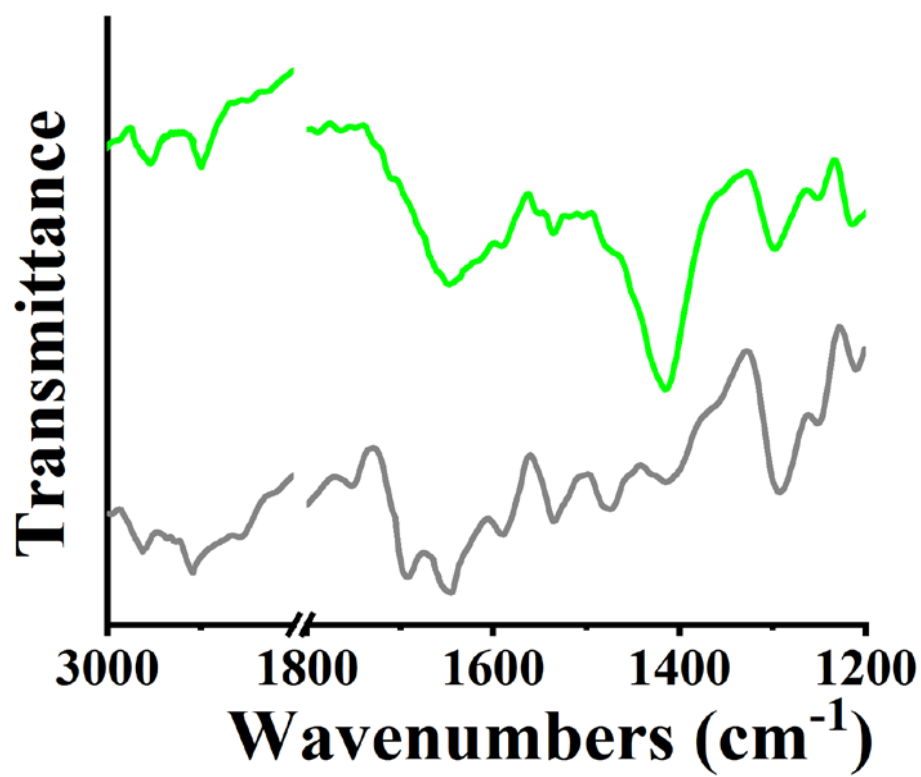

**Figure S1.** FT-IR spectra of BOC-modified MoS$_2$ nanosheets **2** (grey) and amino-modified MoS$_2$ nanosheets **3** (green).

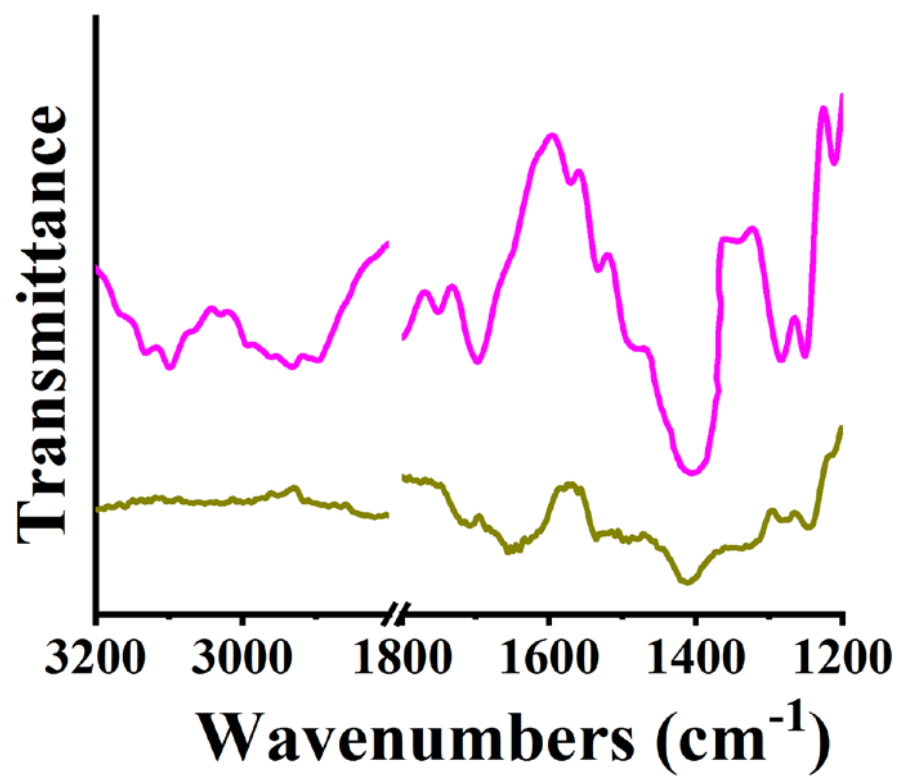

**Figure S2.** FT-IR spectra of oxidized CNHs **5** (pink) and pristine CNHs (yellow).

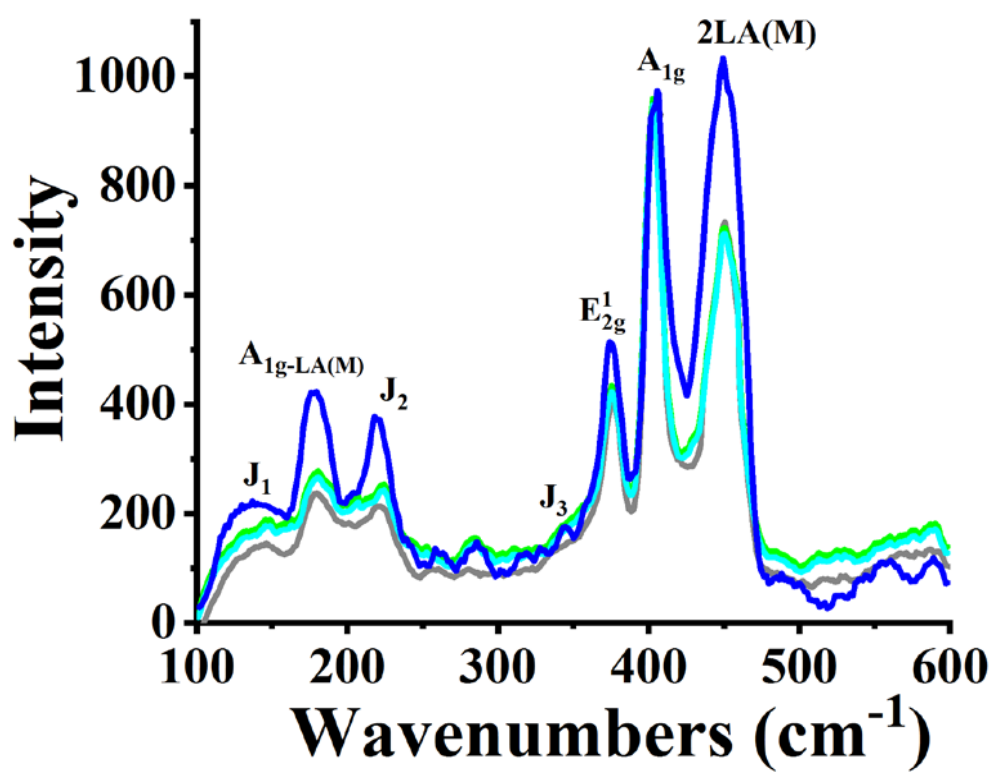

**Figure S3.** Raman spectra of exfoliated 1T-MoS$_2$ (blue) and modified MoS$_2$ materials **2** (grey), **3** (green) and **4** (cyan), recorded upon excitation at 633 nm.

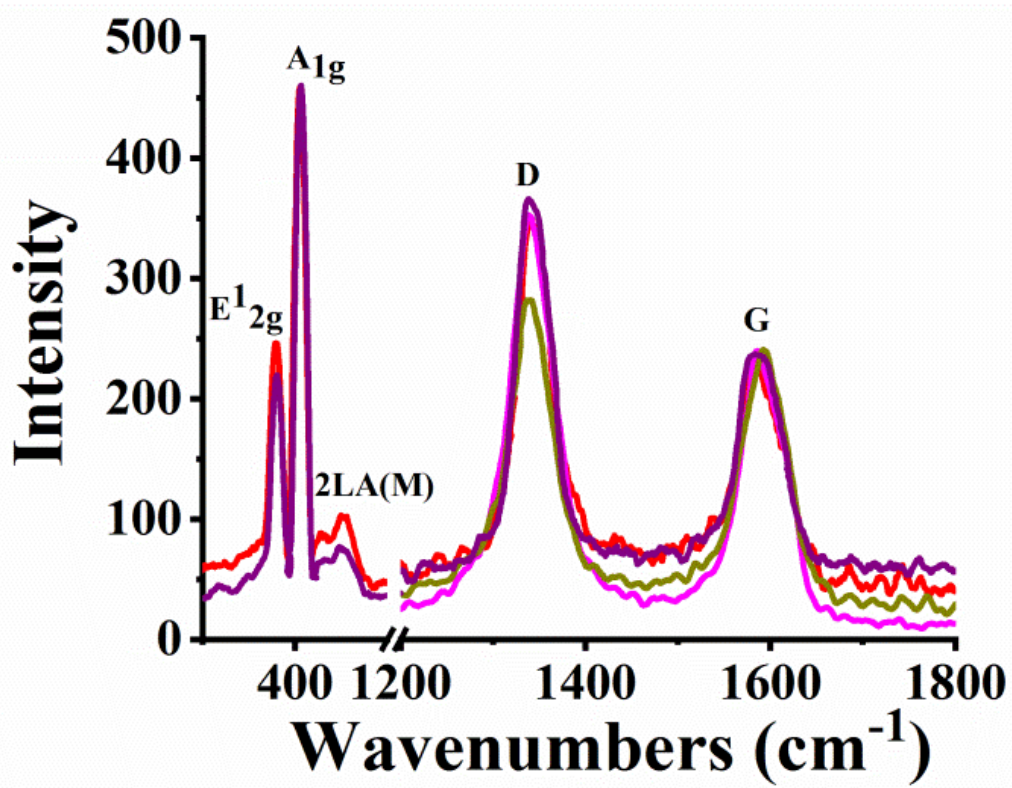

**Figure S4.** Raman spectra of pristine CNHs (dark yellow), oxidized CNHs **5** (pink), **6** (black) and heterostructure **8** (purple) recorded upon excitation at 514 nm.

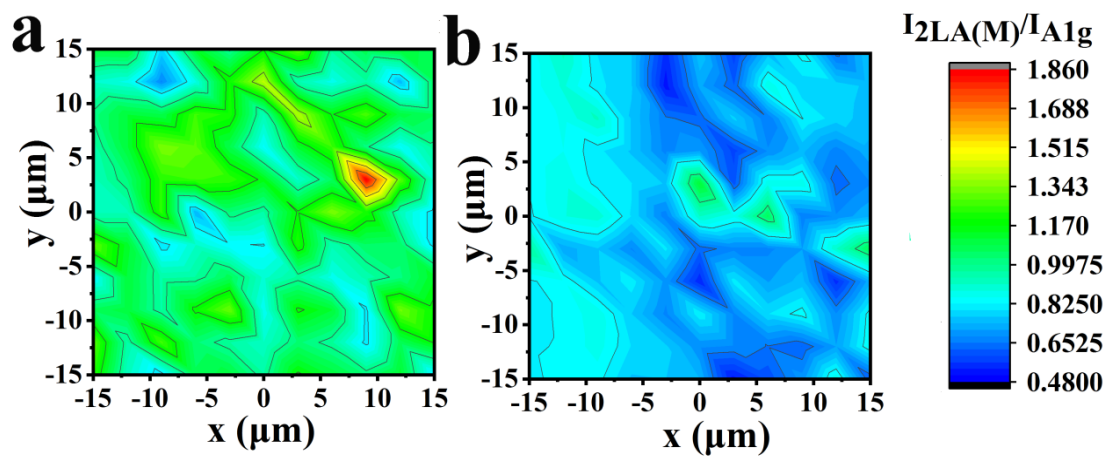

**Figure S5.** Raman maps of (a) exfoliated 1T-MoS$_2$ and (b) modified MoS$_2$ material **4** recorded upon excitation at 633 nm.

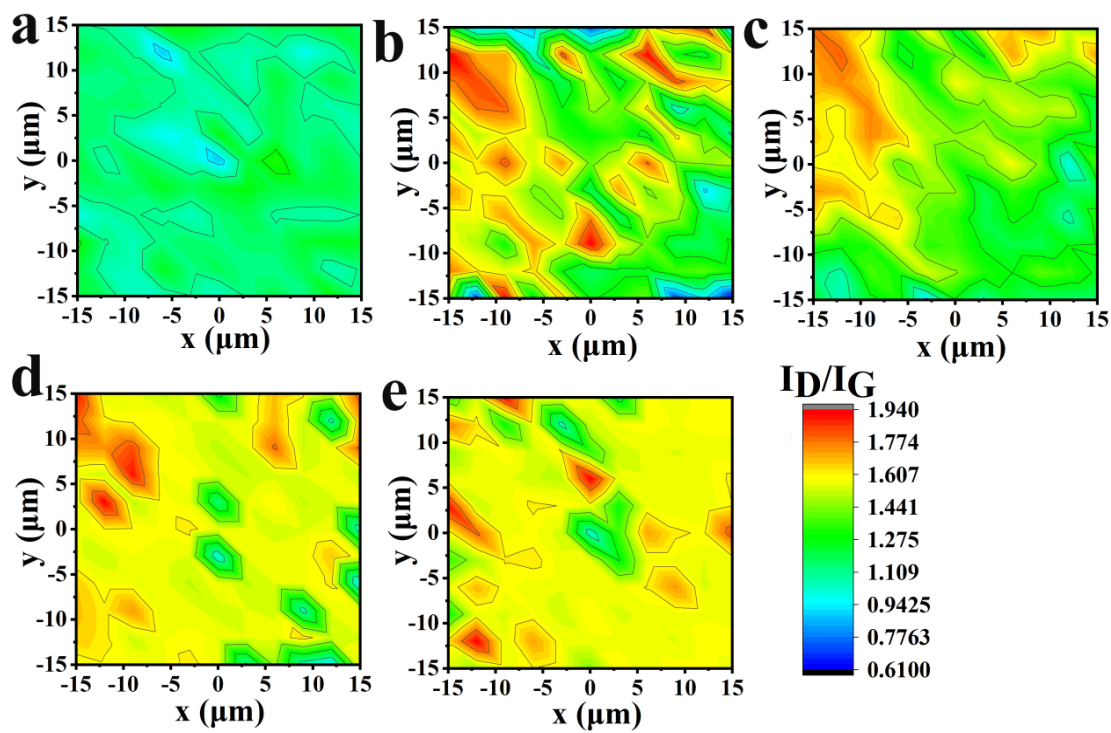

**Figure S6.** Raman maps for (a) pristine CNHs, (b) oxidized CNHs **5**, (c) **6**, (d) **7** and (e) **8** recorded upon excitation at 514 nm.

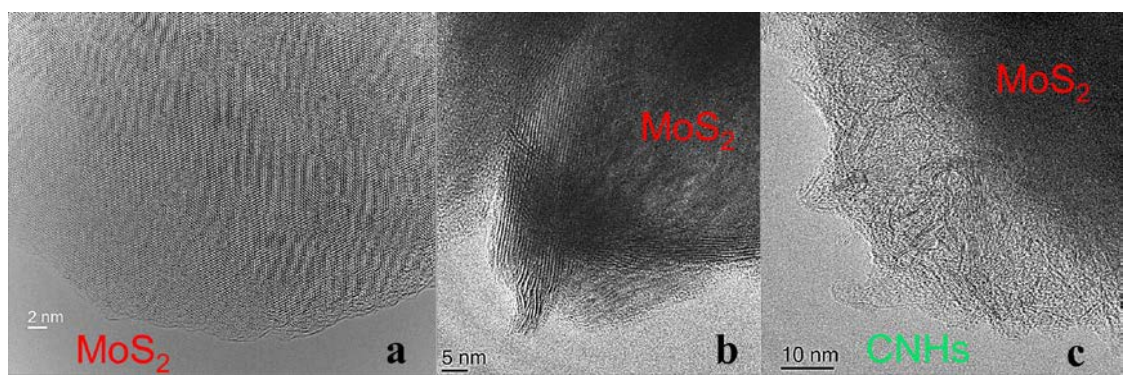

**Figure S7.** HRTEM image of (a, b) exfoliated MoS$_2$ flake, and (c) heterostructure **8** showing the presence of CNHs and MoS$_2$.

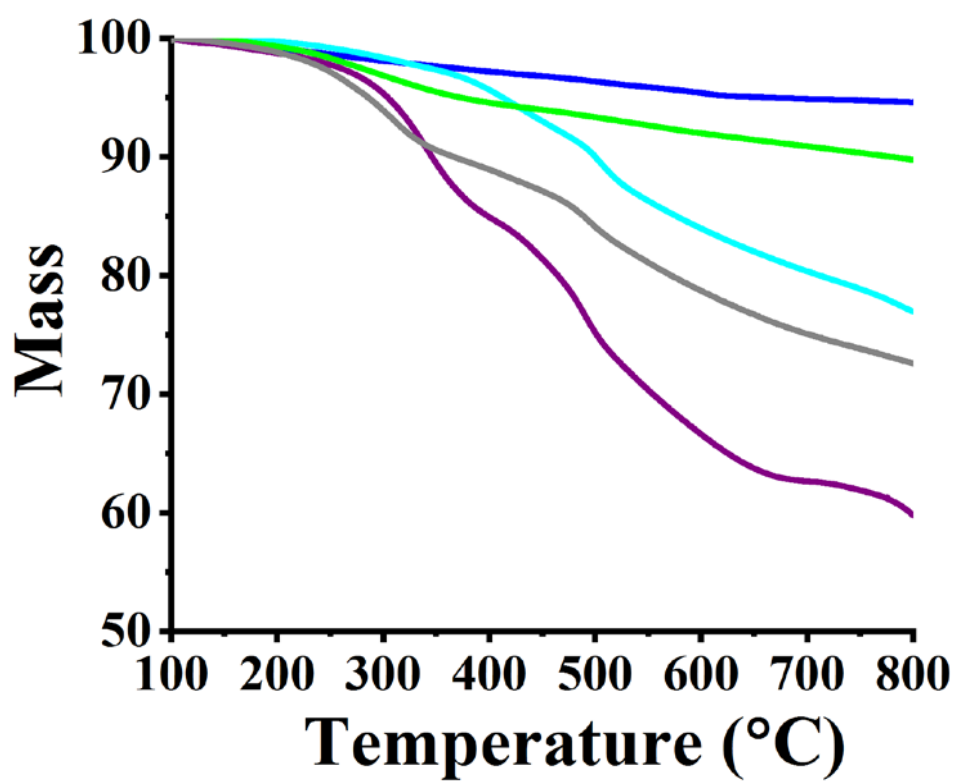

**Figure S8.** TGA graphs for exfoliated 1T-MoS$_2$ (blue), BOC-modified MoS$_2$ material **2** (grey), amino-modified MoS$_2$ material **3** (green), modified MoS$_2$ material **4** (cyan) and heterostructure **8** (purple), recorded under nitrogen atmosphere.

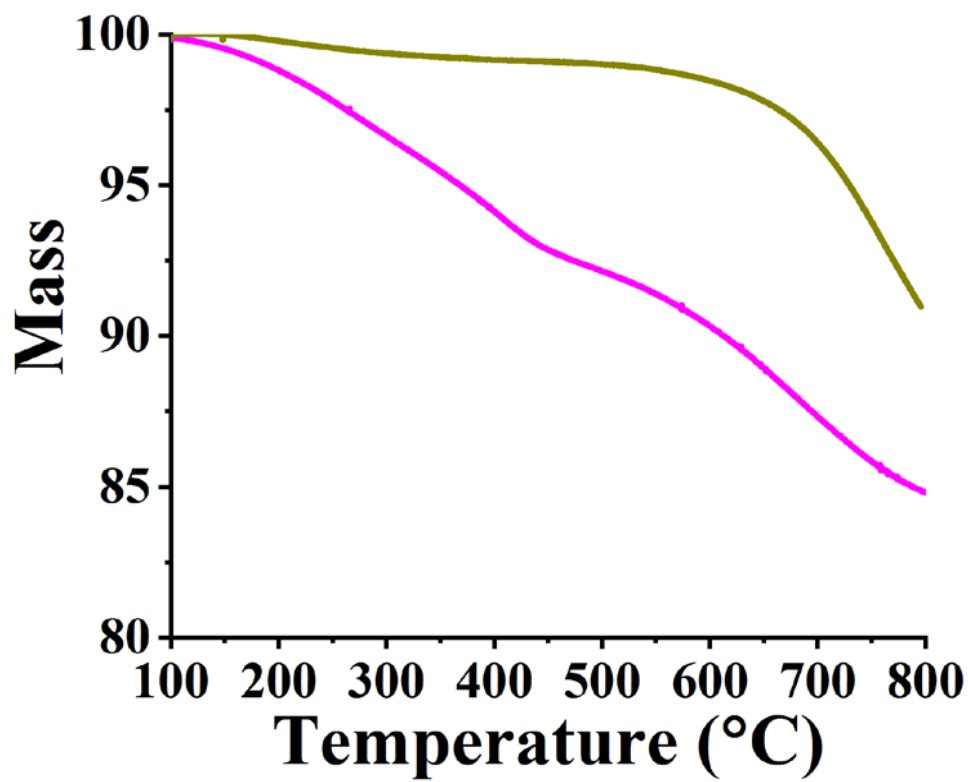

**Figure S9.** TGA graphs for pristine CNHs (yellow) and oxidized CNHs **5** (pink) recorded under nitrogen atmosphere.

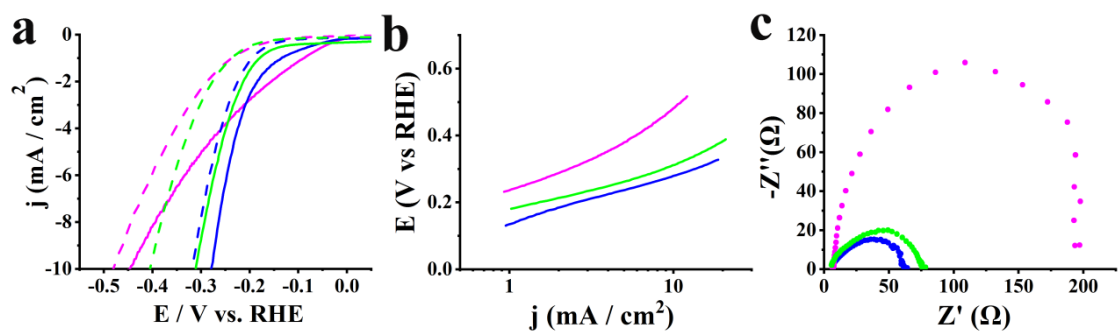

**Figure S10.** (a) LSVs for HER before (solid lines) and after 10,000 cycles (dashed lines), (b) Tafel slopes and (c) Nyquist plots, for exfoliated 1T-$MoS_2$ (blue), material **3** (green) and oxidized CNHs **5** (pink). LSVs obtained at 1,600 rpm rotation speed and 5 mV/s scan rate in aqueous 0.5 M $H_2SO_4$.

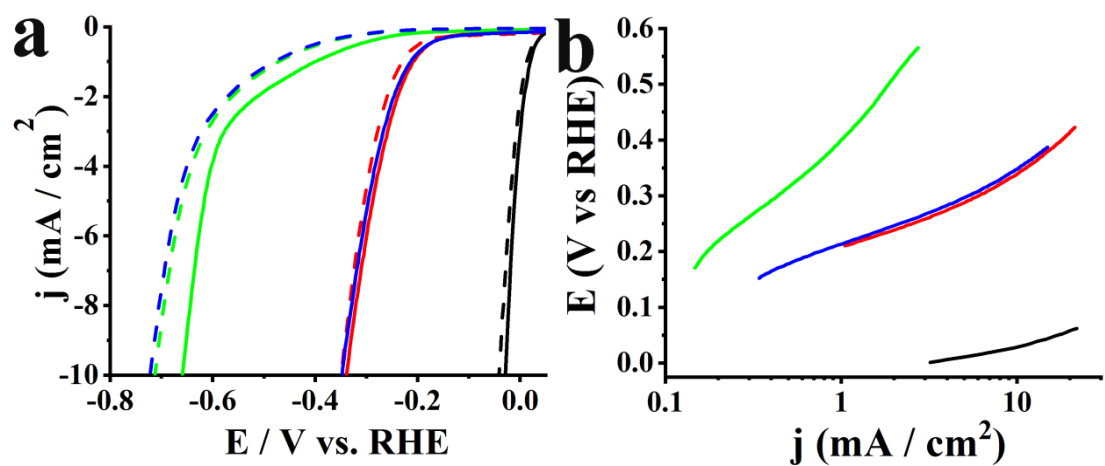

**Figure S11.** (a) LSVs for HER before (solid lines) and after 10,000 cycles (dashed lines) and (b) Tafel slopes for mixtures of **3** and **5** (green), **4** and **6** (blue) and materials **6** (red), **9** (black). LSVs obtained at 1,600 rpm rotation speed and 5 mV/s scan rate in aqueous 0.5 M $H_2SO_4$.